\documentclass[12pt,a4paper]{iopart}
\usepackage{bm}
\usepackage{epsfig}
\usepackage{iopams}
\usepackage{amssymb}
\usepackage{color}

\newcommand{\rlj}[1]{{#1}}

\newcommand{\ee}{\mathrm{e}}

\newcommand{\WW}{\mathbb{W}}
\newcommand{\TT}{\mathbb{T}}
\newcommand{\PP}{\mathbb{P}}

\newcommand{\ZZ}{\mathcal{Z}}

\newcommand{\eps}{\varepsilon}

\begin{document}

\title{Absence of dissipation in trajectory ensembles biased by currents}
\author{Robert L. Jack}
\address{Department of Physics, University of Bath, Bath BA2 7AY, UK}
\author{R. M. L. Evans}
\address{School of Mathematics, University of Leeds, Leeds LS2 9JT, UK}

\begin{abstract}
We consider biased ensembles of trajectories associated with large deviations of currents in equilibrium systems.
The biased ensembles are characterised by non-zero currents and lack the time-reversal symmetry of the equilibrium state,
 but we show that they retain a generalised time-reversal symmetry, involving a spatial transformation that inverts the current.  This means
 that these ensembles lack dissipation.  Hence, they differ significantly from non-equilibrium steady states where currents
 are induced by external forces.  
{
One consequence of this result is that maximum entropy assumptions (MaxEnt/MaxCal), widely used for modelling thermal systems away from equilibrium, have quite unexpected implications, including apparent superfluid behaviour in a classical model of shear flow.
}
\end{abstract}

\maketitle

\section{Introduction}

The mathematical theory of large deviations underlies the rigorous formulation
of equilibrium statistical mechanics and thermodynamics~\cite{Touchette2009,Ruelle}, showing how the free
energy of a very large system is related to the probability of certain rare fluctuations in that system.
In addition to the familiar canonical and microcanonical ensembles used in that context, large
deviation theory can also be applied to \emph{ensembles of trajectories}~\cite{Bod-Der,JL}.  One considers a physical
system evolving in time: for long trajectories, ergodicity implies that time-averaged quantities almost always converge
to their equilibrium averages.  Nevertheless, by focussing attention on 
the rare trajectories in which this convergence does not occur, large deviation theory can reveal unexpected behaviour.
Examples include fluctuation theorems~\cite{gallavotti,jarz97,kurchan98,maes99,crooks2000,LebSpohn99}, and the existence of dynamical phase transitions in both
non-equilibrium systems and supercooled liquids~\cite{lecomte05,lecomte07,garrahan07,garrahan09,hedges09}.  
In the context of sheared systems, it has also been proposed that
rare trajectories of an equilibrium system can be used to predict its response to shear, beyond the linear-response 
regime~\cite{evans04,evans05,evans08,evans14}.

To study these rare trajectories, it is useful to define new ensembles of trajectories via biases (or constraints) on the dynamical evolution of the original system, so that
\emph{typical} trajectories within the new ensembles correspond to the \emph{rare events} of interest in the original system.
In this work, we concentrate on the case where the original system is at equilibrium, and the rare trajectories of interest are those
where a time-averaged current has an atypical (non-zero) value.  (Here, a current is a generic observable that is odd under time-reversal.  Equilibrium states are time-reversal symmetric, so average currents all vanish at equilibrium.)  

We highlight a symmetry of these biased ensembles, which implies that
while they support anomalous currents, they do so without dissipation.  
%In this sense, trajectories from these biased ensembles differ
%qualitatively from those found in systems driven by external forces.   
Motivated by previous work on sheared systems~\cite{evans04,evans05,evans08,evans10},
we illustrate our results using a schematic model of a sheared
fluid, so the relevant current is the shear rate.  However, we frame our main argument in terms of a fairly general Hamiltonian system in contact with a heat bath, and we consider a general class of currents.  
The presence of the heat bath is not central to the argument, but it is useful in clarifying some parts of the argument, especially
when considering the response of the system to non-conservative external forces.

Briefly, our main result is that the rare trajectories that realise a particular current
$J$ are related by time-reversal to the trajectories that realise current $-J$.  This forbids the flow of dissipative currents such as the flow
of heat into the bath, since the direction of such currents must be invariant under $J\to-J$.  It follows that trajectories of
systems sheared by external forces are generically different from the rare large-shear trajectories obtained by large-deviation theory.
The role of time-reversal symmetry and of currents in this argument means that our results are related to previous work by Maes
and co-workers in the context of non-equilibrium response theory~\cite{maes06,maes09}.  The main consequence of our result is that we identify 
qualitative differences between responses to external forces on a system, and dynamical biases (or constraints) on time-averaged currents.

%The symmetry relation that we discuss is linked to time-reversibility of equilibrium states, and is concerned with the fluctuations
%of `currents', by which we mean generic observables that 
%are odd under time-reversal.  Fluctuations of time-reversal symmetric and antisymmetric quantities have been discussed by , and our results are related to that work~.  
%{ 
%Here we investigate the physical properties of biased ensembles of trajectories and demonstrate the differences from driven systems, both in general and in some models of fluids.
%}

The biased ensembles that we consider are also identical \cite{evans05} to those obtained by Jaynes's maximum entropy inference (MaxEnt) prescription applied to trajectories and using current as a macroscopic observable (in which case it is also known as MaxCal \cite{MaxCal1,MaxCal2,MaxCal3}). Hence our results demonstrate that MaxEnt/MaxCal does  {\em not} yield the non-equilibrium dynamics of driven systems, contrary to the widely-held hypothesis \cite{MaxCal1,MaxCal2,MaxCal3,evans05}.

\section{General setting}

We consider a system that evolves in time under a dynamics with
some stochastic element.  We use $x$ to indicate a generic configuration (or phase space point).
We concentrate on cases where $x=(\vec{q},\vec{p})$, with $\vec{q}$ being a vector of generalized co-ordinates
and $\vec{p}$ a vector of conjugate momenta.  However, the results may be easily generalised to other models such as
Markov chains, where $x$ would represent an element of a discrete configuration space.

\subsection{Equilbrium dynamics}

We first define an equilibrium dynamics and an associated energy function $E(x)$.  We fix Boltzmann's constant $k_{\rm B}=1$.
We use $X$ to indicate a trajectory of the system, running from an initial time $t=-\tau$ to a final time $t=\tau$. %
We write $(X)_t = x(t)$ for the state of the system at time $t$.  By ``an equilbrium dynamics'', we mean (i) that the system's dynamical rules
have the Boltzman distribution $p(x) \propto \ee^{-E(x)/T}$ as a steady state, and (ii) that this steady state is time-reversal symmetric.  
(We further assume that the steady state is unique.)
An example is the case where
$x=(\vec{q},\vec{p})$, the energy is 
\begin{equation}
E = \sum_i \frac12 p_i^2 + V(\vec{q}) ,
\end{equation}
and the system evoves by Langevin equations
\begin{eqnarray}
\partial_t q_i & =  & p_i  \, ,
 \\
\partial_t p_i & = & - \frac{\partial V}{\partial q_i }- \lambda_i p_i + \sqrt{2\lambda_i T} \eta_i  \, .
\label{equ:dpdt}
\end{eqnarray}
%{\NEW Rob, I have replaced $\lambda$ by $\lambda_i$.}
Here, $\lambda_i$ is a friction constant and $\vec{\eta}$ a vector of white noises with mean zero and 
$\langle \eta_i(t) \eta_j(t') \rangle = \delta_{ij} \delta(t-t')$.  For $\lambda_i=0$ we recover Hamiltonian time evolution.
We emphasise that the equilibrium steady state associated with this evolution is time-reversal symmetric for all $\lambda_i$, as may be demonstrated
explicitly by writing an appropriate Fokker-Planck equation (see~\ref{sec:operators}, below).  

It is useful to define an operator $\TT$ which gives the time-reversed counterpart of a trajectory $X$.
The momenta $p_i$ are odd under time-reversal, so  
we can write 
\begin{equation} (\TT X)_t= \overline{x}(-t), \end{equation} 
where $\overline{x}=(\vec{q},-\vec{p})$
is obtained by reversing all momenta in configuration $x$.  (The overbar should not be confused with any kind of average.)
Considering these dynamics and working in the steady state of the system,
one may define a probability density $P_{\rm eq}(X)$ over all possible dynamical trajectories. 
This distribution has a time-reversal symmetry:
\begin{equation}
P_{\rm eq}(X) = P_{\rm eq}(\TT X) .
\label{equ:T-eq}
\end{equation}

\subsection{\label{sec:driv}Driven dynamics}

Next we define a dynamics where the system is driven out of equilibrium by some external forces.  That is, we modify
(\ref{equ:dpdt}) to
\begin{equation}
\partial_t p_i = - \frac{\partial V}{\partial q_i }- \lambda_i p_i + f_i + \sqrt{2\lambda_i T} \eta_i
\label{equ:eom-neq}
\end{equation}
where the external forces $f_i$ may be collected into a vector $\vec{f}$.  These forces
 are assumed to be non-conservative, that is, they cannot
be obtained as the gradient of any external potential.  
The probability distribution for trajectories in the steady state of this non-equilibrium dynamics is denoted by
$P_{\rm neq}(X)$.  Due to the non-conservative forces, there is no time-reversal symmetry: 
$P_{\rm neq}(X) \neq P_{\rm neq}(\TT X)$.

In this work, we are interested in cases where the external forces $\vec{f}$ break a spatial reflection symmetry
of some kind.  For example, one might have $E(\vec{q},\vec{p})=E(-\vec{q},-\vec{p})$ so that the system's equilibrium behaviour
is unchanged if all co-ordinates are inverted.  More generally, define an operator $\PP$ by 
\begin{equation}(\PP X)_t=\tilde{x}(t),\end{equation}
where $\tilde{x}$ is related to $x$ through inversion of one or more co-ordinates (and their conjugate momenta).
We assume that the equilibrium dynamics are invariant under this transformation, in which case 
\begin{equation} P_{\rm eq}(X) = P_{\rm eq}(\PP X). \end{equation}
However, we further assume that the external forces $\vec{f}$ break this symmetry so that $P_{\rm neq}(X) \neq P_{\rm neq}(\PP X)$.

Note that the driven dynamics considered here is different from the ``driven dynamics'' of~\cite{evans04,evans05,evans08,evans14,evans10,chetrite14}.  We consider here a general non-equilibrium driving force, where they consider a specific force that is chosen so that to mimic particular rare events in the original system.

\subsection{Biased dynamics}

The non-conservative external forces $\vec{f}$ in the driven dynamics will
induce currents within the system.  
We define an instantaneous current $j=j(x)$.  The dependence of $j(x)$ on $x$ can be fairly
general but in order to be interpreted as a current, 
we require it changes sign under time-reversal: $j(\overline{x})=-j(x)$.  
A simple case (see below) is that
$\vec{f}$ corresponds to a shear stress, in which case the associated current would be a strain rate.  The external
forces break the spatial reflection symmetry $\PP$, and
we also assume that $j$ changes sign under this inversion: $j(\tilde{x})=-j(x)$.
The total current associated with a trajectory $X$ is
\begin{equation}
J(X) = \int_{-\tau}^\tau j(x(t)) \mathrm{d}t.
\label{equ:J-int}
\end{equation}
From the symmetry properties of the current, one has $J(\TT X)=-J(X)$ and $J(\PP X)=-J(X)$.

Following~\cite{LebSpohn99,Bod-Der,evans05}, we are concerned here 
with the large deviations of $J$ in the limit $\tau\to\infty$.  To this end we define a biased ensemble of trajectories
\begin{equation}
P_{\rm bias}(X|\nu) =  P_{\rm eq}(X) \cdot \frac{\ee^{\nu J(X)} }{ \ZZ(\nu) }
\label{equ:bias}
\end{equation}
where $\nu$ is the strength of the bias, and $\ZZ$ is a normalisation constant (or dynamical partition sum).
For a physical interpretation of this ensemble, note that for large $\tau$, the ensemble $P_{\rm bias}$ is very close (in a precise sense~\cite{touchette13}) to
the ensemble of
trajectories obtained by constraining the total current $J$ to some particular value.  That is, the biased distribution $P_{\rm bias}$
gives the \emph{least unlikely} trajectories that are consistent with a particular ($\nu$-dependent) value of the total
current $J$.
{
Alternatively, (\ref{equ:bias}) is the ensemble with maximum combinatorial entropy relative to $P_{\rm eq}$, subject to a conditioning on the average current $J$. This is exactly the ensemble that results from Jaynes' MaxEnt or MaxCal procedure~\cite{MaxCal1,MaxCal2}.
}

Given all these definitions, one easily sees that the probability of a time-reversed trajectory $\TT X$ in the biased ensemble is equal
to the probability of the original trajectory $X$ in an ensemble with the opposite bias: that is,
\begin{equation}
P_{\rm bias}(\TT X|\nu) = \frac{ P_{\rm eq}(\TT X) \ee^{\nu J(\TT X)} }{ \ZZ(\nu) }
= \frac{ P_{\rm eq}(X) \ee^{-\nu J(X)} }{ \ZZ(\nu) } = P_{\rm bias}(X|-\nu) .
\end{equation}
Similarly one finds that $P_{\rm bias}(\PP X|\nu) = P_{\rm bias}(X|-\nu)$.  Hence, substituting $X\to \TT X$, one has a ``generalised time-reversal''
symmetry for the biased ensemble:
\begin{equation}
P_{\rm bias}(\PP \TT X|\nu) =  P_{\rm bias}(X|\nu) .  
\label{equ:PT-bias}
\end{equation}
That is, given a trajectory $X$, one may obtain another trajectory with equal probability by first inverting the direction of time
and then inverting those co-ordinates associated with the operator $\PP$. See also~\cite{maes06}.
\ref{sec:operators} illustrates these considerations further, using an operator representation.

We emphasise that there is typically no counterpart of (\ref{equ:PT-bias}) for the driven ensemble $P_{\rm neq}$.  In the following,
we will show that (\ref{equ:PT-bias}) means that the biased ensemble is free from dissipation, while the driven ensemble $P_{\rm neq}$
typically corresponds to a physical dissipative process. 
We also note that we have assumed so far that the system has non-zero frictional and noise forces (i.e., $\lambda_i>0$),  so that its steady state is an
equilibrium Boltzmann-distributed state at temperature $T$.  However, the analysis leading to (\ref{equ:PT-bias}) holds also
for purely deterministic (Hamiltonian) dynamics with $\lambda_i=0$.  The noise and damping forces are useful here since they ensure
that the driven system (with $f\neq0$) eventually converges to a steady state.

We now demonstrate the resulting differences between the driven and biased ensembles in various examples, both close to and far from equilibrium.

\section{Illustrative examples}
\subsection{Linear response}

{
We first illustrate these differences by considering linear response to the bias $\nu$ and the force $f$.  
}
For any observable
$O(t)$, we work at equilibrium ($\nu=0=f$) and calculate a derivative with respect to $\nu$
 (see for example~\cite{garrahan09}).  The result is
\begin{equation}
\frac{\mathrm{d}}{\mathrm{d}\nu} \langle O(0) \rangle_{\rm bias} = \int_{-\infty}^\infty \mathrm{d}t \langle O(0) j(t) \rangle_{\rm eq}
\label{equ:dOdnu}
\end{equation}
Similarly, if we take a force $f$ conjugate to the current $j$, then~\cite{hansen-mac}
\begin{equation}
\frac{\mathrm{d}}{\mathrm{d}f} \langle O(0) \rangle_{\rm neq} = \frac1T \int_{0}^\infty \mathrm{d}t \langle O(0) j(t) \rangle_{\rm eq} .
\label{equ:dOdsig}
\end{equation}
Clearly if the correlation function $\langle O(0) j(t) \rangle_{\rm eq}$ is even in time then $\frac{\mathrm{d}}{\mathrm{d}\nu} \langle O(0) \rangle_{\rm bias}
= 2T\frac{\mathrm{d}}{\mathrm{d}f} \langle O(0) \rangle_{\rm bias}$.  On the other hand, if the correlation function is odd
then $\frac{\mathrm{d}}{\mathrm{d}\nu} \langle O(0) \rangle_{\rm bias}=0$ while the response to $f$ may be finite.

The simplest case is $O(t)=j(t)$, in which case we measure the response of the current $j$ itself.  The relevant correlation function is even, so the responses
differ by a factor of $2T$.  On the other hand, if $O(t)$ depends only on the rotor orientations $\theta(t)$ then \rlj{$\langle O(0) j(t)\rangle_{\rm eq}$} is odd
in time.  (To see this, note that trajectories at equilibrium have probabilities equal to their time-reversed counterparts, and that if $O$ is a function
of only of positions $q_i$ then the integral
$\int_{-\infty}^{\infty}\mathrm{d}t\, O(0) j(t)$ 
changes sign under time-reversal of any trajectory.  So all contributions to the correlation function vanish when considering
pairs of trajectories related by time reversal.)  

Hence, for such observables, there is no response to the bias $\nu$, {while there is typically a finite response to the force $f$.}
These results are generic in ensembles of the form considered here, provided that the driving force used in the non-equilibrium
ensemble is conjugate to the current $j$, so that (\ref{equ:dOdsig}) holds.  These results will be useful in section \ref{RobsModel}.

\subsection{\label{MikesModel}A continuously sheared fluid}

\begin{figure}
\hspace{1cm}\includegraphics[width=14cm]{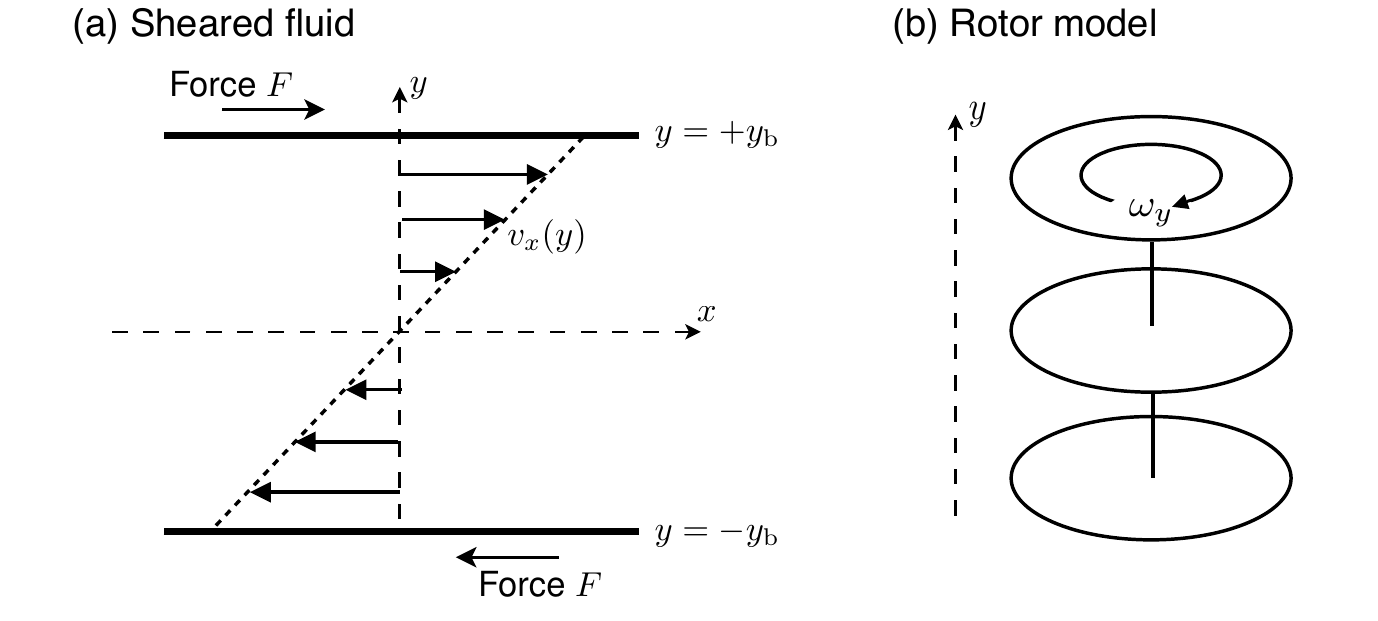}
\caption{(a)~Schematic of a sheared system between two parallel plates at $y=\pm y_{\rm b}$, with forces $F$ and $-F$ applied
to the top and bottom rates.  The mean velocity at height $y$ is $v_x(y)$ with $v_x(y)=y\dot{\gamma}$ in a state of uniform shear rate.  
We imagine periodic boundaries in the $x$ and $z$ directions. (b)~Simplified `rotor' model, consisting of a set of discs placed along the $y$-axis. The angular velocity of the disc at position $y$ is $\omega_y$, which is analogous to the velocity $v_x(y)$ in (a).}
\label{fig:shear_simple}
\end{figure}

{
We now illustrate the abstract definitions of the different ensembles by a commonplace example.
}
Fig.~\ref{fig:shear_simple}(a) illustrates  a sheared system, for which biased ensembles of the form (\ref{equ:bias}) were discussed in~\cite{evans05,evans08a}.
A slab of fluid sits between two parallel walls, at $y=\pm y_{\rm b}$, with periodic
boundaries in the $x$ and $z$ directions. (There should be no confusion between these Cartesian co-ordinates and the 
notation $x$ for a generic phase space point.) 
Forces are applied to the plates and the system (eventually) converges to a steady state with a finite shear rate.
In this steady state, the external forces are constantly injecting work into the system, this energy acts to heat up the fluid, and 
eventally flows out through the walls of the system, which we assume to be maintained at constant temperature $T$ by some external
thermostat.

The particles within the fluid evolve according to Hamiltonian's equations, except for particles close to the boundary, where they feel (stochastic) thermal noise forces,
and shear forces associated with the parallel plates.  The equations of motion for the particle momenta 
can be written in the form (\ref{equ:eom-neq}), except that the thermal noise forces $\eta_i$, damping forces $\lambda_i \omega_i$,
and external forces $f_i$ act only at the boundary.  
In the absence of external forces, one has a time-reversal symmetric steady state.  

On introducing a shear stress $\sigma$, a shear rate in the system can be defined as $\dot\gamma=(v_x(y_{\rm b}) - v_x(-y_{\rm b}))/2y_{
\rm b}$ where $v_x(y)$
is the average of the $x$-component of the velocity of the fluid, within a thin slab at height $y$.  This shear rate will correspond
to the general `current observable' of the previous section: $j=\dot\gamma$.  
It is a linear combination of velocities, so is manifestly odd under time-reversal,
as required.
The total shear $\gamma$ is then easily obtained by a time integral, and we identify the time-integrated current
$J=\gamma=\int_{-\tau}^\tau j(t) \mathrm{d}t$, as in (\ref{equ:J-int}).  
To apply our general discussion to this system, 
the relevant spatial inversion symmetry $\PP$ is the co-ordinate transform $(x,y,z)\to(-x,y,z)$.  The equilibrium dynamics are invariant
under this transformation; operation with $\PP$ inverts the
velocities $v_x$ so it also takes $j\to -j$ as required.

It follows that the generalised time reversal symmetry (\ref{equ:PT-bias}) holds for this system. 
That is, the biased ensemble of trajectories (\ref{equ:bias}) for 
this model is invariant under time-reversal followed by a spatial reflection in the plane $x=0$.

To see the connection of this result to dissipation, we compare this ensemble with a driven (sheared) steady state.  
In the driven system, one expects currents of energy to flow through the
system: the work done by external forces injects energy into the system, this energy flows into the microscopic degrees of freedom of the fluid, and eventually
leaves the system as heat, via the external boundaries.  If one could reverse the arrow of time, these dissipative energy currents would be reversed: heat would flow
into the fluid at the boundaries and appear to perform work on the external plates.  A subsequent spatial reflection through $x=0$ does
not reverse the direction of these energy currents.  Thus, the driven steady state (with finite shear stress) does not respect the symmetry (\ref{equ:PT-bias}).

If follows that the dissipative energy currents that naturally flow in driven systems are inconsistent with the symmetry relation Eq.~(\ref{equ:PT-bias}),
so they are forbidden within the biased ensemble (\ref{equ:bias}).
This is the sense in which biased ensembles such as (\ref{equ:bias})
differ from driven non-equilibrium ensembles in which external forces act at the boundaries.

\subsection{A model sheared system\label{sec:average-forces}}
%{\NEW (New section)}

To make these arguments concrete,
we analyse a simple model system in which Eq.~(\ref{equ:PT-bias}) has important consequences.   
We consider a set of $N$ rotors (similar to that in~\cite{evans14}), each with moment of inertia $I$, as illustrated in Fig.~\ref{fig:shear_simple}(b).
We draw an analogy between the rotor velocity $\omega_y$ and
the velocity $v_x(y)$ for the sheared system shown schematically in Fig.~\ref{fig:shear_simple}. This one-dimensional set of rotors can then be regarded as a highly simplified model of the interactions within a sheared fluid.

\newcommand{\ft}{f_{\rm t}}
\newcommand{\fb}{f_{\rm b}}
\newcommand{\etat}{\eta_{\rm t}}
\newcommand{\etab}{\eta_{\rm b}}

In analogy with the interparticle forces in a classical fluid, rotors apply purely conservative torques $u'(\Delta\theta)=\varepsilon\sin(\Delta\theta)$ to their neighbours, that depend only on the relative angle $\Delta\theta_i\equiv\theta_{i+1}-\theta_i$.
To model the application of shear stress and heat on the boundaries of the fluid, we apply an additional external torque $\ft(t)$ to the topmost rotor, and $\fb(t)$ to the rotor at the bottom.  The equations of motion are
\begin{eqnarray}
\label{systemEq}
	I \partial_t \omega_1 &=& u'(\Delta\theta_1) + \fb(t)  \nonumber \\
	I \partial_t \omega_2 &=&  u'(\Delta\theta_2) - u'(\Delta\theta_1)  \nonumber \\
	\ldots & &   \nonumber \\
	I \partial_t \omega_i &=&  u'(\Delta\theta_i) - u'(\Delta\theta_{i-1})  \nonumber \\
	\ldots & &   \nonumber \\
	I \partial_t \omega_N &=& -u'(\Delta\theta_{N-1}) + \ft(t).
\end{eqnarray}
These equations fully specify the properties of the rotors.  The boundary forces $f_{\rm t,b}$ follow from properties of the thermal bath to which the rotors are coupled.  
They have both deterministic and stochastic parts, arising  from applied macroscopic shear stress and heat exchange respectively. We write
\begin{eqnarray*}
	\ft &=& \lambda_0\Omega-\lambda_0\omega_N + \etat(t) \sqrt{2\lambda_0 T},  \\
	\fb &=& -\lambda_0\Omega-\lambda_0\omega_1 + \etab(t) \sqrt{2\lambda_0 T}, 
\end{eqnarray*}
where $\lambda_0$ is a friction coefficient associated with the dissipative coupling to the boundary, $\lambda_0\Omega$ is the external torque on the system, and $\etat(t)$ and $\etab(t)$ are independent random noises, with coefficients chosen to respect the Einstein relation for a heat bath of temperature $T$.

At equilibrium ($\Omega=0$), the $\eta_{\rm t,b}$ are the usual Gaussian noises, and the system is time-reversal symmetric. 
In the driven case, $\Omega$ is non-zero, while the noises have the same form as at equilibrium. In that case, work is done on the system by the applied torques at the boundaries, which leads to average shear flow.  At the same time, heat energy (in the form of disordered motion) flows to the boundaries where it is dissipated.  The system will converge to a steady state in which these energy fluxes balance.

In the biased case, no explicit driving force is applied, so $\Omega=0$, but Eq.~\ref{equ:bias} means that the noise from the heat bath is sampled non-uniformly, so that the stochastic functions $\eta_{\rm t,b}(t)$ can acquire non-zero expectation values, which induce shear flow. On the face of it,  one might imagine that $\langle\etat\rangle$ in the biased ensemble plays the same role as $\Omega\sqrt{\lambda_0/2T}$ in the driven ensemble, in which case the biased and driven ensembles would be similar. In fact, the two ensembles behave very differently, as we shall now see.

Whatever the ensemble, the mean (time-averaged) torque applied at the top boundary is 
\(
	\langle \ft \rangle = \langle I \partial_t \omega_N + u'(\Delta\theta_{N-1}) \rangle
\)
and, since $\langle\partial_t\omega_N\rangle=0$ in a steady state, we have
\begin{equation}
		\langle \ft \rangle =  \varepsilon\langle\sin(\Delta\theta_{N-1}) \rangle
		\label{equ:ft-sintheta}
\end{equation}
for any steady-state ensemble. (There is also a similar expression for $\langle \fb\rangle$.)
%
%{ 
This means that  the mean torque on the boundary can be obtained from the ($i$-dependent) distribution $P(\Delta\theta_i)$ of relative angles between neighbouring rotors.
To make progress, we define the symmetry operation $\PP$ as the co-ordinate transformation $(\theta_i)\to(-\theta_i)$, which has the properties specified in Sec.~\ref{sec:driv}.  
Also note that reversing the arrow of time leaves $P(\Delta\theta_i)$ unchanged, while the symmetry operation $\PP$ changes the sign of $\Delta\theta$.  Hence the combined $\PP\TT$ operation transforms $P(\Delta\theta_i)$ to $P(-\Delta\theta_i)$.  From~(\ref{equ:PT-bias}), the biased ensemble is invariant under $\PP\TT$ so $P(\Delta\theta_i)=P(-\Delta\theta_i)$ within this ensemble (for all $i$).  That is, the distribution of $\Delta\theta_i$ is symmetric in the biased ensemble, so $\langle\sin\Delta\theta_i\rangle_{\rm bias}=0$.
%}

From here, the startling implication of (\ref{equ:ft-sintheta}) is that the mean applied torque on the boundary must vanish, $\langle \ft\rangle=0$ in the biased ensemble, thus describing a thermodynamic system induced to flow (shear) continuously by the application of no mean force at all.    The system, in the biased ensemble, thus behaves like a superfluid, which is not consistent with the responses of classical systems to external driving.

{
\subsection{\label{RobsModel}A sheared model with internal noise}

Our final example is a modified version of the above model, similar to those considered in~\cite{evans15,evans10}.  
In contrast to the previous section, all the rotors are coupled to the thermal bath.  
%
%We next consider an alternative version of the above model, with stochastic dynamics. That is, the noise and friction are now internal to the system, as opposed to being applied only at the boundary as was the case above. A stochastic rotor model of this form was
%discussed in~\cite{evans15} and, with a more elaborate form of conservative potential, in~\cite{evans10}.
%}
For the purposes of this work it is sufficient to consider a system with
just three rotors -- this very simple system is already sufficient to illustrate the symmetry (\ref{equ:PT-bias}) of biased ensembles, and the breaking of
this symmetry in driven systems.  It is also simple enough that numerical results are easy to obtain.

 As before, the co-ordinates of the system are the angles $\theta_1,\theta_2,\theta_3$ which specify the orientation of the rotors.
Each rotor has moment of inertia $I$ so the momenta in the system are $I\omega_i$ with $\omega = \dot \theta_i$.  
The energy of the system is 
\begin{equation}
E=\sum_i \frac12 I \omega_i^2 + u(\theta_1-\theta_2) + u(\theta_2-\theta_3) 
\end{equation}
with $u(\Delta\theta)=-\eps\cos\Delta \theta$.  Frictional forces act on the velocity differences between all rotors, and
a constant driving torque of strength $\sigma$ is applied to the boundary rotors, so that
the equations of motion are
\begin{eqnarray}
I \partial_t \omega_1 = -u'(\theta_1-\theta_2) - \lambda(\omega_1-\omega_2) + \sqrt{2\lambda T} \eta_{1} - \sigma
\nonumber \\ 
I \partial_t \omega_2 = -u'(\theta_2-\theta_3) - u'(\theta_2-\theta_1)  - \lambda(2\omega_2-\omega_1-\omega_3) +\sqrt{2\lambda T} (\eta_{2} - \eta_1)
\nonumber \\ 
I \partial_t \omega_3 = -u'(\theta_3-\theta_2) - \lambda(\omega_3-\omega_2) - \sqrt{2\lambda T}  \eta_{2} + \sigma
\label{equ:eom-rotor}
\end{eqnarray}
where $u'(\Delta\theta)=\eps\sin\Delta \theta$ is the derivative of $u$, 
and $\eta_{1,2}$ are uncorrelated Gaussian noises with mean zero and variances $\langle \eta_i(t) \eta_j(t')\rangle = \delta_{ij}\delta(t-t')$,
as above.  Clearly $\partial_t(\omega_1+\omega_2+\omega_3)=0$ so
we fix the global momentum to zero without loss of generality: $\omega_1+\omega_2+\omega_3=0$.

For $\sigma=0$ one has an equilibrium state with time-reversal symmetry.  The system is also invariant under inversion of all positions and momenta:
that is, the symmetry operation $\PP$ is defined by taking $\tilde{x}=(-\vec{\theta},-\vec{\omega})$.
The shear rate is $j=(\omega_3-\omega_1)/2$ which is odd under both time-reversal and under $\PP$. (The factor of $2$ comes from the linear extent
of the system along the $y$-direction, for a system of $N$ rotors one would have $j=(\omega_N-\omega_1)/(N-1)$.) Thus, defining a biased ensemble according
to (\ref{equ:bias}) with $J=\frac12 \int_{-\tau}^\tau (\omega_3-\omega_1)\mathrm{d}t$, the generalised time-reversal symmetry 
(\ref{equ:PT-bias}) applies in this system.

The behaviour of the model is controlled by three dimensionless parameters.  The first two of these are $\eps/T$ and $\sigma/T$, 
which set the strength of the conservative
forces and the external forces, respectively.  The final parameter is $\lambda_0 = \lambda/\sqrt{IT}$ which sets the strength of
the damping.  The rotor co-ordinates $\theta$ are naturally dimensionless so it remains only to fix a time unit.  There are several intrinsic time scales
within the system: we focus on $\tau_0 = I/\lambda$, which is equal to the velocity relaxation time in the weak-force limit $\eps/T\to0$.  When showing
numerical results we use units such that $\tau_0=1$.  This time
scale is natural for systems with intermediate damping strength and moderate values of $\eps/T$.  Other time scales are more
relevant for very strong damping ($\tau_{\rm B}=\lambda/T=\tau_0\lambda_0^2$); for very weak damping ($\tau_{\rm th}=\sqrt{I/T}=\tau_0\lambda_0$);
or very strong conservative forces ($\tau_{\rm harm}=\sqrt{I/\eps}=\tau_0\lambda_0\sqrt{T/\eps}$).

\subsubsection{Structure in sheared states}

\begin{figure}
\hfill 
\includegraphics[width=10cm]{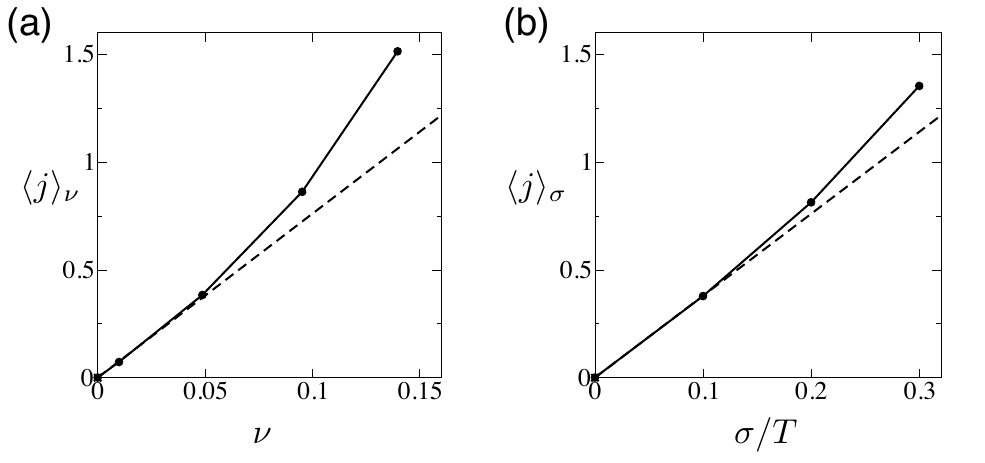} \hspace{2cm}
\caption{%(a)~Illustration of a simple one-dimensional rotor model, here with $N=3$.  The rotors are arranged along the $y$-direction, with
%the angular velocity $\omega_y=\omega(y)$ corresponding to the velocity $v_x(y)$ in the simple shear geometry of
%Fig.~\ref{fig:shear_simple}.  { Interactions via equal and opposite torques exist between nearest neighbour rotors only.}
Dependence of the normalised shear rate 
$\langle j \rangle = \langle \omega_3 -\omega_1 \rangle/2$ on bias $\nu$
(a) and applied torque $\sigma$ (b).  The unit of time is $\tau_0=1$.  The dashed lines are linear response results, obtained 
by numerical evaluation
of the correlation functions in (\ref{equ:dOdnu},\ref{equ:dOdsig}). Expanding about the equilibrium state, 
one has $\frac{\mathrm{d}}{\mathrm{d}\nu} \langle j(0) \rangle_{\rm bias}
=2T\frac{\mathrm{d}}{\mathrm{d}\sigma} \langle O(0) \rangle_{\rm neq}$.}
\label{fig:rotor-jdep}
\end{figure}

We analyse this model using numerical simulation, in two cases: (i) a non-equilibrium ensemble which depends on the driving
force $\sigma$; and (ii) the biased ensemble (\ref{equ:bias}) which depends on the bias strength $\nu$.  We consider only
the case where $\eps/T=1$ and $\lambda_0=0.3$, which is a representative state point that is sufficient to illustrate our
main results.  For equilibrium simulations
and for case (i), we use solve the equations of motion by the method of Bussi and Parrinello~\cite{bussi}, as described in~\cite{crooksLangevin}.  
The time step is fixed at $0.01\tau_0$.
For biased ensembles, we use the same 
scheme in conjunction with transition path
sampling methods~\cite{TPS}, which are natural tools for sampling ensembles of the form of (\ref{equ:bias}), see for example~\cite{hedges09,speck12,fullerton13}.  
We consider
trajectories of length $2\tau$ with $\tau=15\tau_0$, which provides a balance between convergence of the large-$\tau$ limit (as required
for studies of large deviations),
and manageable computational cost.  Note also that the symmetry relation (\ref{equ:PT-bias}) applies for all $\tau$, not only
in the large-$\tau$ limit. However, the biased ensemble can be identified with a steady state only when $\tau$ is large~\cite{garrahan09,chetrite14}.

  Fig.~\ref{fig:rotor-jdep} shows
how the shear rate $\langle j \rangle$ depends on the applied bias $\nu$ and applied force $\sigma$.
To investigate the structure
of the system at finite shear rate, we measure the distribution of the angular difference $\Delta\theta=(\theta_2-\theta_1)\,\rm{modulo}\,2\pi$.  At equilibrium
$P(\Delta\theta) \propto \ee^{\eps\cos\Delta\theta/T}$, consistent with the Boltzmann distribution.

\begin{figure}
\includegraphics[width=\columnwidth]{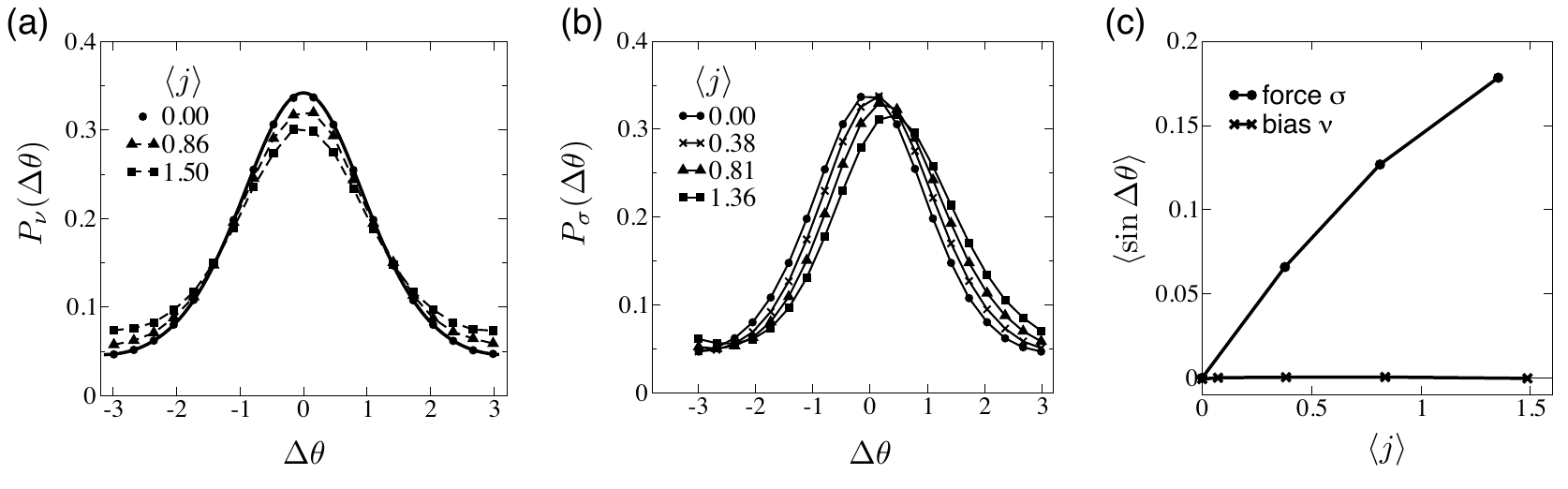}
\caption{(a) Distributions of the angular difference $\Delta\theta=(\theta_2-\theta_1)\,\rm{mod}\,2\pi$ in biased ensembles with 
$0\leq \langle j\rangle < 1.5$,
as labelled.  The solid line is $P(\Delta\theta)\propto \ee^{(\eps/T)\cos\theta}$, dashed lines are guides to the eye. 
(b) Distributions of $\Delta\theta$ for driven ensembles ($\sigma>0$), over a similar range of $\langle j\rangle$ (all lines
are guides to the eye).
As discussed in the main text,
the distribution in the biased ensemble is symmetric while the distribution in the driven ensemble is not.
(c) Mean conservative force $\langle\sin\Delta\theta\rangle$ plotted parametrically as a function of the current, in both biased
and driven ensembles.  In the biased case, the symmetry of $P(\Delta\theta)$ means that the average force is always zero.
}
\label{fig:dhist}
\end{figure}

Fig.~\ref{fig:dhist} shows
corresponding distributions for the biased and driven ensembles, over comparable ranges of the  shear rate $\langle j \rangle$.  The distributions differ qualitatively:
for the biased state, $P(\Delta\theta)$ is a symmetric function of $\Delta\theta$ while for the driven state, this symmetry is lacking. 
To further accentuate
this difference, we consider the mean force between the rotors $\eps\langle \sin(\theta_2-\theta_1)\rangle$.  For a direct comparison,
we plot the mean force parametrically against the shear rate $\langle j \rangle$.  The force is positive in the driven
ensemble but vanishes in the biased ensemble, consistent with the symmetry of $P(\Delta\theta)$,
%{
that was responsible for the vanishing of mean torque discussed in section \ref{MikesModel}.
Since the symmetry of $P(\Delta\theta)$ in the biased case follows from (\ref{equ:PT-bias}), the numerical results in Fig.~\ref{fig:dhist} illustrate the effect of
this generalised time-reversal symmetry. The driven system ($\sigma>0$) lacks the symmetry (\ref{equ:PT-bias}), as is clear from the asymmetry of $P(\Delta\theta)$ in Fig.~\ref{fig:dhist}(b).
%}

\subsubsection{Relation to dissipation}

To illustrate the relation of these results to dissipation, we note that the conservative part of the torque applied to the second rotor by the first is 
$-u'(\theta_2-\theta_1)$, so the first rotor does work on the second at a rate $\dot{Q}_{12}=-\omega_2 u'(\theta_2-\theta_1)$.  
 We can interpret
this as an energy current from rotor 1 to rotor 2.  The $\PP$ operation inverts both $u'$ and $\omega_2$, leaving the energy current
invariant, but the $\TT$ operation inverts $\omega_2$ only.  Thus, the combination $\PP\TT$ changes the sign of $\dot{Q}_{12}$, 
and (\ref{equ:PT-bias}) implies $\langle \dot{Q}_{12} \rangle_{\nu}=0$ in the biased ensemble.  However, for the driven ensemble, we have generically
$\langle \dot{Q}_{12} \rangle_{\sigma}>0$. Note this quantity is positive, independent of the sign of $\sigma$: the sign of the dissipation
is independent of the direction of the applied force, as expected.

Fig~\ref{fig:q_vs_j} shows numerical results for $\dot{Q}_{12}$, plotted parametrically as a function of the shear rate.  As expected there is no dissipative
current in the biased ensemble.  In the driven ensemble, the outer rotors do work on the central one: this energy is then dissipated through friction, 
maintaining the steady state.
[We note in passing that since the system is in a steady state, we have $\partial_t \langle \cos(\theta_2-\theta_1)\rangle=0$ even for $\sigma>0$, and hence $\langle (\omega_1-\omega_2)\sin(\theta_2-\theta_1)\rangle=0$.  Hence one always has  $\langle\omega_1\sin(\theta_2-\theta_1)\rangle=\langle\omega_2\sin(\theta_2-\theta_1)\rangle$, the question is whether these two quantities vanish individually, or not.]

\begin{figure}
\hspace{2cm}\includegraphics[width=7cm]{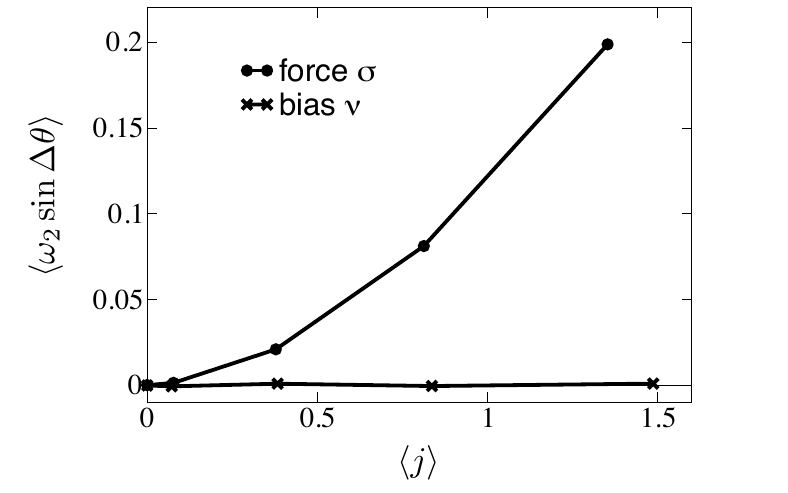}
\caption{Average energy current $\dot{Q}_{12}= \langle \omega_2 \sin(\theta_2-\theta_1)\rangle$, comparing biased ($\nu>0)$ and driven  $(\sigma>0$)
ensembles.  The current is plotted parametrically against the shear rate $\langle j \rangle$.   The numerical results are consistent with the absence
of dissipation in the biased ensemble.
In the driven ensemble, note that this current is unchanged by the spatial
transform $\PP$ so it is an even function of $\sigma$, and 
$d\dot{Q}_{12}/d\langle j \rangle=0$ at $\langle j \rangle=0$.  This contrasts with the mean force shown in Fig.~\ref{fig:dhist}c, which changes its
sign under $\PP$, and is an odd function of $\langle j \rangle$.
}
\label{fig:q_vs_j}
\end{figure}

\subsubsection{Force balance and non-zero stochastic forces}
\label{sec:force}

Finally, it is instructive to take the average of Eq.~\ref{equ:eom-rotor} in the biased ensemble, to make contact with Sec.~\ref{sec:average-forces}.
For the first rotor we obtain
\begin{equation}
0 = \eps \langle  \sin(\theta_2-\theta_1) \rangle_{\rm bias} + \lambda \langle \omega_2- \omega_1 \rangle_{\rm bias} + \sqrt{2\lambda T}\langle \eta_1 \rangle_{\rm bias}
\label{equ:3rotor-ave}
\end{equation}
For $\nu>0$ then clearly $\langle \omega_2- \omega_1 \rangle_{\rm bias}>0$, but as noted above, $\langle  \sin(\theta_2-\theta_1) \rangle_{\rm bias}=0$. The sum of the last two terms on the right hand side of (\ref{equ:3rotor-ave}) is analogous to the average force $\langle\ft\rangle_{\rm bias}$ in Sec.~\ref{sec:average-forces}, and this average force is zero, as noted in that section.  Since $\langle \omega_2- \omega_1 \rangle_{\rm bias}>0$, 
it must therefore be that the noise term has a non-zero average within the biased ensemble
\begin{equation}
\langle \eta_1 \rangle_{\rm bias} = -\sqrt{\frac{\lambda}{2T}} \langle \omega_2- \omega_1 \rangle_{\rm bias} .
\label{equ:ave-eta}
\end{equation}
Thus, as noted in Sec.~\ref{sec:average-forces}, the finite shear rate that appears in the biased ensemble is sustained by a finite value for a thermal noise force, due
to the presence of the bias.

\section{Conclusion}

The main result of this work is Eq.~(\ref{equ:PT-bias}), which is a symmetry of biased ensembles of trajectories.  Our discussion shows
that this symmetry places significant constraints on the behaviour that can be observed in these ensembles.  In particular, there is a class
of \emph{protected observables} whose average value is always zero, even when currents are flowing in the system.  These protected observables are related to dissipative
processes in the system, and we argue that their absence means that biased ensembles are non-dissipative.  This behaviour is in contrast to
that found in systems that are driven away from equilibrium by external forces.

Ensembles of trajectories of the form of (\ref{equ:bias}) appear naturally in calculations based on maximum-entropy inference, since they provide the most
likely (or least unlikely) trajectories that are consistent with constraints that are applied to time-integrated currents~\cite{evans05}.  
{
Thus, for dissipative non-equilibrium systems, our findings invalidate the popular MaxCal procedure if it is conditioned on a current.
}

From a physical
perspective, it is not clear to us \emph{why} the most likely trajectories in biased ensembles should be free from dissipation. This is a consequence of the time-reversal
symmetry of the equilibrium state that survives even in these far-from-equilibrium biased ensembles.  We hope that further work on the properties
of large deviations in non-equilibrium systems might lead to insights in this direction.  For example, the absence of dissipation is related to the response
theory of~\cite{maes09} and might also be connected to the effective interactions that arise in biased ensembles~\cite{jackSollich15}.

\ack We thank Peter Sollich for helpful discussions.  RLJ was supported by the EPSRC through grant EP/I003797/1.

\begin{appendix}
\section{Operator representations of the generalised time-reversal symmetry}
\label{sec:operators}

As discussed in~\cite{LebSpohn99,lecomte05,lecomte07}, biased ensembles of the form (\ref{equ:bias}) are related to ``tilted'' generators
or master operators.  The symmetry (\ref{equ:PT-bias}) has a simple interpretation in terms of these operators.  We give a brief
discussion of this interpretation here (an alternative approach based on path integrals and action functionals can also be used to obtain
similar results~\cite{maes06,maes09}).

\newcommand{\TTh}{\hat{\TT}}
\newcommand{\PPh}{\hat{\PP}}
\newcommand{\pih}{\hat{\pi}}

Our starting point is the master operator (the adjoint of the generator) of the equilibrium stochastic process of interest.  
To analyse the case given in (\ref{equ:dpdt}), we introduce
a representation of the phase space of the system based on Dirac kets $|x\rangle$.  The probability distribution $P(x)$ for system's phase 
space point corresponds to a ket $|P\rangle = \int\mathrm{d}x P(x)|x\rangle$ which
evolves in time according to $\partial_t |P\rangle = \WW_{\rm eq} |P\rangle$ with~\cite{vanKampen}
\begin{equation}
\WW_{\rm eq} = \sum_i \left[ -p_i \frac{\partial}{\partial q_i} +  \left(\frac{\partial E}{\partial q_i}+\lambda p_i\right) \frac{\partial}{\partial p_i} 
 + \lambda  \left( 1 + T \frac{\partial^2}{\partial p_i^2} \right) \right]
 \label{equ:WWeq}
\end{equation}
Applying this operator to the equilibrium (Boltzmann) distribution yields $\WW_{\rm eq}|P_{\rm eq}\rangle=0$, confirming that
this is indeed the steady state of the model.
To analyse the time-reversal symmetry of this model, we introduce an operator $\TTh$ which inverts the direction of momenta:  
$\TTh|x\rangle=|\overline{x}\rangle$.  We introduce a
second operator $\pih$ which is diagonal, with elements $\ee^{-E(x)/T}$.  The time-reversal symmetry of the equilibrium
ensemble of trajectories (\ref{equ:T-eq}) corresponds to the operator equation
\begin{equation}
\WW_{\rm eq}^\dag =  (\TTh\pih)^{-1} \WW_{\rm eq} (\TTh \pih)
\label{equ:TT-op}
\end{equation}
This equation may be verified directly from the definitions of the various operators.
(Note that $\TTh^{-1}=\TTh$, since it simply corresponds to a reversal of momenta.)
We also introduce an operator $\PPh$ corresponding to the spatial transformation $\PP$, by taking $\PPh|x\rangle=|\tilde{x}\rangle$.
If the dynamics is invariant under $\PP$, one has an operator equation 
\begin{equation}
\PP \WW_{\rm eq} \PP=\WW_{\rm eq}.
\label{equ:PP-op}
\end{equation}
(For the operator $\WW_{\rm eq}$ in 
(\ref{equ:WWeq}), this relationship is easily verified as long as $\partial E/\partial q_i$ is odd in $q_i$ for those co-ordinates $q_i$ which are inverted by $\PP$.)
We also note that $\int\mathrm{d}x \langle x|\WW_{\rm eq}|P\rangle=0$, independent of $|P\rangle$: this corresponds to conservation
of probability.  

To analyse the driven ensemble, we write $\WW_{\rm neq} = \WW_{\rm eq} - \sum_i f_i \frac{\partial}{\partial p_i} $ where the $f_i$ are
the external forces (assumed independent of $p_i$).  Since these forces are non-conservative, the relation (\ref{equ:TT-op}) does not apply.
However, the relation $\int\mathrm{d}x \langle x|\WW_{\rm neq}|P\rangle=0$ holds also for this
non-equilbrium dynamics, since probability is (of course) still conserved.

To analyse the biased ensemble, we write $\WW_{\rm bias}(\nu) = \WW_{\rm eq} + \nu \hat{j}$ where the operator $\hat{j}$ is diagonal
with elements $j(x)$.  The theory associated with this operator is discussed in~\cite{LebSpohn99,lecomte07,jackSollich10}.  The operator $\WW_{\rm bias}(\nu)$ does
not have a probability-conservation property $\int\mathrm{d}x \langle x|\WW_{\rm bias}(\nu)|P\rangle\neq0$.  However,
the steady state probability distribution of $x$ in the biased ensemble is controlled by the largest eigenvalue of $\WW_{\rm bias}$ and the
associated left and right eigenvectors.  Given the properties of the current discussed above (it is odd under both $\TT$ and $\PP$), then we have
 $\PPh \hat{j} \PPh=-\hat{j}=\TTh \hat{j} \TTh$.  We also have
$\pih^{-1} \hat{j} \pih=\hat{j}$, since these operators are all diagonal.  Hence it follows from (\ref{equ:TT-op}) that
\begin{equation}
\WW_{\rm bias}(\nu)^\dag =  (\PPh\TTh\pih)^{-1} \WW_{\rm bias}(\nu) (\PPh\TTh \pih)
\label{equ:PT-op}
\end{equation}
which is the promised operator equation corresponding to the symmetry (\ref{equ:PT-bias}).

To see the consequences of this equation, suppose that $\langle L|$ is the dominant
left eigenvector of $\WW(\nu)$ so that $|L\rangle$ is the dominant right eigenvector of $\WW(\nu)^\dag$.  Then from (\ref{equ:PT-op}) the dominant right
eigenvector of $\WW(\nu)$ is $|R\rangle =  (\PPh \TTh \pi)|L\rangle$.  The probability of configuration $x$ in the steady state is 
$P_{\rm bias}(x|\nu)\propto\langle L|x\rangle\langle x|R\rangle$~\cite{jackSollich10}, so that
$P_{\rm bias}(x|\nu)\propto L(x) L(\tilde{\overline{x}}) \pi(x)$ where $\tilde{\overline{x}}$ is the phase space point obtained by applying $\TT\PP$.  Hence
\begin{equation}
P_{\rm bias}(\tilde{\overline{x}}|\nu) = P_{\rm bias}(x|\nu)
\label{equ:pbias-sym}
\end{equation}
which is the symmetry relation for the steady state distribution of the biased process.  Averages of one-time observables in the biased
ensemble are fully determined by $P_{\rm bias}(x|\nu)$, so (\ref{equ:pbias-sym}) specifies which quantities can have non-zero
values in that ensemble, and which are constrained equal to zero by symmetry.

The strength of this operator approach is that the same algebraic structure can hold for a variety of different models.  For example, there
are many discrete Markov chain models where symmetries of the form (\ref{equ:PT-op}) apply, including the simple symmetric exclusion process (SSEP)
biased by the total current~\cite{Bod-Der,appert08,jack15hyper}.  Thus, while we have concentrated throughout on systems with continuous co-ordinates $x=(\vec{q},\vec{p})$,
the operator formalism allows straightforward generalisations to overdamped Langevin dynamics (where $x=\vec{q}$) or to Markov chains such as the SSEP.

\end{appendix}

\end{document}